\documentclass[a4paper,11pt]{article}
\usepackage{pos}
\usepackage{physics}
\usepackage{tabularx}
\usepackage{color}

\title{Investigations of decuplet baryons from meson-baryon interactions in the HAL QCD method}
\ShortTitle{Decuplet baryons from meson-baryon interactions in the HAL QCD method}

\manuallySeparateAuthors

\author*[a,b]{Kotaro Murakami}
\author[a,b]{, Yutaro Akahoshi}
\author[a]{, Sinya Aoki}
\author[c]{ and Kenji Sasaki}

\author{\\for HAL QCD Collaboration}

\affiliation[a]{Center for Gravitational Physics, 
Yukawa Institute for Theoretical Physics, Kyoto University, Kitashirakawa Oiwakecho, Sakyo-ku, 
Kyoto 606-8502, Japan}

\affiliation[b]{RIKEN Nishina Center, Saitama 351-0198, Japan}

\affiliation[c]{Division of Scientific Information and Public Policy (SiPP), Center for Infectious Disease Education and Research (CiDER), Osaka University, 2-8 Yamadaoka, Suita City, Osaka 565-0871, Japan}

\emailAdd{kotaro.murakami@yukawa.kyoto-u.ac.jp}
\emailAdd{yutaro.akahoshi@yukawa.kyoto-u.ac.jp}
\emailAdd{saoki@yukawa.kyoto-u.ac.jp}
\emailAdd{kenjis@cider.osaka-u.ac.jp}

\abstract{We study decuplet baryons from meson-baryon interactions, in particular, $\Delta$ and $\Omega$ baryons from P-wave $I=3/2$ $N\pi$ and $I=0$ $\Xi\bar{K}$ interactions, respectively. The interaction potentials are calculated in the HAL QCD method using 3-quark-type source operators at $m_{\pi} \approx 410~\textrm{MeV}$. We use the conventional stochastic estimation of all-to-all propagators combined with the all-mode averaging to reduce statistical fluctuations. We have found that two potentials have quite similar behaviors, suggesting that a mass difference between $\Delta$ and $\Omega$ comes mainly from a difference of kinematical structure between $N\pi$ and $\Xi \bar K$, rather than their interactions. The scattering phase shifts calculated from the potentials indicate that $\Delta$ and $\Omega$ baryons exist as bound states in this lattice setup, whose binding energies are consistent with those obtained from 2-point functions.}

\FullConference{%
 The 38th International Symposium on Lattice Field Theory, LATTICE2021
  26th-30th July, 2021
  Zoom/Gather@Massachusetts Institute of Technology
}

\begin{document}
\raisebox{80pt}[0pt][0pt]{\hspace*{92mm} YITP-21-145, RIKEN-QHP-509}
\vspace{-9mm}
\maketitle

\section{Introduction}
Most of hadrons can be understood well in the quark model, while some exceptions, called exotic hadrons, have been found in various experiments. Exploring properties or internal structures of such exotic hadrons from QCD is one of the biggest issues in hadron physics. Since exotic hadrons typically appear as resonances due to non-perturbative QCD interactions, studies in lattice QCD are mandatory to understand them.

In lattice QCD, masses of hadron resonances may be estimated from corresponding 2-point functions, while decay rates are evaluated only from hadron scatterings. The HAL QCD method~\cite{Ishii:2006ec, Aoki:2009ji, Ishii:2012ssm} is suitable to analyze hadron scatterings in lattice QCD, where we extract hadron interaction potentials from NBS wave functions and extract scattering phase shifts by solving the Schor\"{o}dinger equations with potentials. Compared with others, this method has an advantage that systems including baryons can be handled more reliably.

As a first step toward studies of exotic hadrons including pentaquarks, we study decuplet baryons, spin $3/2$  baryons symmetric under quark flavor exchanges. All decuplet baryons except $\Omega$ are resonances: For example, $\Delta$ baryon decays into $N\pi$. On the other hand, $\Omega$ is a stable particle and appears as a bound state of $\Xi\bar{K}$. Thus we expect to see in the HAL QCD method whether the $N\pi$ interaction for the resonant $\Delta$ is different from the  $\Xi\bar{K}$ interaction for the bound $\Omega$. In our study, we employ $u$ and $d$ quark masses closed to the $s$ quark mass, where the SU(3) flavor symmetry is slightly broken, to explore an origin of the difference between two baryons. In this setup, $\Delta$ baryon exists as a stable particle.

This paper is organized as follows. In Sec.~\ref{sec:intro_of_HAL}, we briefly review the HAL QCD method in meson-baryon systems. In Sec.~\ref{sec:setups}, we show our setup of the analysis for $N\pi$ and $\Xi\bar{K}$ interactions. In Sec.~\ref{sec:results}, we present numerical results for $N\pi$ and $\Xi\bar{K}$ potentials, scattering phase shifts and binding energies. Sec.~\ref{sec:conclusion} is devoted to a conclusion of this paper.

\section{HAL QCD method in meson-baryon systems}\label{sec:intro_of_HAL}

In this paper, we employ the time-dependent HAL QCD method~\cite{Ishii:2012ssm}, which improves the original HAL QCD method~\cite{Ishii:2006ec, Aoki:2009ji} for an extraction of potentials in lattice QCD calculations. In this method, we use R-correlator defined as
\begin{eqnarray}
R_{\alpha}(\vb{r},t) 
= \frac{F_{\alpha}(\vb{r},t)}{C_{M}(t)C_{B}(t)},
\end{eqnarray}
where $C_{M}(t)$ and $C_{B}(t)$ are 2-point correlation functions for meson and baryon, respectively, and $F$ is a 4-point correlation function of the meson-baryon system. Explicitly, we have
\begin{eqnarray}
F_{\alpha}(\vb{r},t) = \bra{0}   M(\vb{r} , t) B_{\alpha}(\vb{0} , t) \ \bar{\mathcal{J}}_{MB} (0) \ket{0},
\end{eqnarray}
where $M(\vb{x},t)$ and $B_{\alpha}(\vb{x},t)$ are meson and baryon sink operators, respectively, and $\bar{\mathcal{J}}_{M B} (t_{0})$ is the source operator which has the same quantum number as that of the meson-baryon system. The R-correlator can be decomposed into elastic and inelastic parts as
\begin{eqnarray}\label{eq:propofRcorr}
R_{\alpha}(\vb{r},t)
 = \sum_{n}   A_{n} \Psi^{W_{n}}_{\alpha}(\vb{r}) \ e^{-\Delta  W_{n}t}+(\textrm{inelastics}),
\end{eqnarray}
where $\Psi^{W_{n}}_{\alpha}$ is the equal-time NBS wave function for the meson-baryon system with energy $W_{n}=\sqrt{k_{n}^2+m_{M}^2}+\sqrt{k_{n}^2+m_{B}^2}$,  $A_{n}$ is a factor independent of $\vb{r}$ and $\alpha$, and $\Delta  W_{n} = W_{n}-m_{M}-m_{B}$ is the energy from the threshold.

The elastic part $A_{n} \Psi^{W_{n}}_{\alpha}(\vb{r})e^{-\Delta  W_{n}t}$ satisfies Schr\"{o}dinger equation as
\begin{eqnarray}\label{eq:scheqofelastic}
 \left( \frac{k^2_{n}}{2\mu} - H_{0} \right) A_{n} \Psi^{W_{n}}_{\alpha}(\vb{r})e^{-\Delta  W_{n}t}
 =\int d^3r' \ U_{\alpha\beta}(\vb{r},\vb{r}') A_{n} \Psi^{W_{n}}_{\beta}(\vb{r'})e^{-\Delta  W_{n}t},
\end{eqnarray}
where $\mu$ is the reduced mass, $H_{0}$ is the free Hamiltonian and $U(\vb{r},\vb{r}')$ is the non-local interaction potential. Noticing that $k_{n}^{2}/2\mu$ can be expanded in terms of $\Delta W_{n}$ as
\begin{eqnarray}
\frac{k_{n}^2}{2\mu} 
= \Delta W_{n} + \frac{1+3\delta}{8\mu}(\Delta W_{n})^2 + \frac{M^2\delta^2}{8\mu}\sum_{k=3}^{\infty} (k+1)\left(\frac{-\Delta W_{n}}{M}\right)^k 
\equiv \sum_{k=1}^{\infty}C^{(k)}_{m_{M},m_{B}}(\Delta W_{n})^{k}, 
\end{eqnarray}
where $M=m_{M}+m_{B}$ and $\delta = (m_{M}-m_{B})/M$, then rewriting $\Delta W_{n}$ in terms of the time derivative, and taking summation over $n$, we obtain
\begin{eqnarray}
\left[\sum_{k=1}^{\infty}C^{(k)}_{m_{M},m_{B}}\left(-\pdv{t}\right)^{k}-H_{0} \right]R_{\alpha}(\vb{r},t)
\simeq \int d^3r' \ U_{\alpha\beta}(\vb{r},\vb{r}')R_{\beta}(\vb{r'},t).
\end{eqnarray}
for a large enough $t$ so as to suppress inelastic contributions. 

According to the Okubo-Marshak expansion~\cite{Okubo:1958} for meson-baryon systems, the leading order term in the derivative expansion of $U_{\alpha\beta}(\vb{r},\vb{r}')$ is given by
\begin{eqnarray}
U_{\alpha\beta}(\vb{r},\vb{r}') \simeq V^{\textrm{LO}}(r)\delta_{\alpha\beta}\delta^{(3)}(\vb{r}-\vb{r}'),
\end{eqnarray}
where $V^{\textrm{LO}}(r)$ can be extracted from $R_{\alpha}(\vb{r},t)$ for any $\alpha$ as
\begin{eqnarray}\label{eq:LOpotential_gen}
V^{\textrm{LO}}(r) = \frac{1}{R_{\alpha}(\vb{r},t)}\left[\sum_{k=1}^{\infty}C^{(k)}_{m_{M},m_{B}}\left(-\pdv{t}\right)^{k}-H_{0} \right]R_{\alpha}(\vb{r},t).
\end{eqnarray}
Since the LO potential is invariant under rotation, we have~\cite{Murano:2013xxa} 
\begin{eqnarray}\label{eq:LOpot_murano}
V^{LO}(r) 
= \frac{\sum_{g\in O}R^{\dag}_{\alpha}(g{\bf r},t)
\left[\sum_{k=1}^{\infty}C^{(k)}_{m_{M},m_{B}}\left(-\pdv{t}\right)^{k}-H_{0} \right]R_{\alpha}(g{\bf r},t)}
{\sum_{g\in O}R^{\dag}_{\alpha}(g{\bf r},t)R_{\alpha}(g{\bf r},t)}, 
\end{eqnarray}
where $O$ is the cubic group. In this paper, we use this equation instead of Eq.~(\ref{eq:LOpotential_gen}) to reduce noises caused by $R_\alpha$ data fluctuating around zero, and restrict an infinite summation over $k$ to $k\le 2$ for $N\pi$ and $k\le 3$ for $\Xi \bar{K}$, respectively, since remaining higher order terms are negligibly small\footnote{There is an alternative method to derive the potentials without the expansion in $\Delta W$ by using at most the 3rd derivatives with respect to the time~\cite{Doi:HALfullyrel}.  We have confirmed that this exact one gives no significant differences from our results
for $N\pi$ and $\Xi \bar K$ potentials, showing that higher order terms are indeed small.}. In addition, we take an average over $\alpha$ to increase statistics.

\section{Setups for the analysis of the $N\pi$ and $\Xi\bar{K}$ interactions}\label{sec:setups}
\subsection{Correlation functions with single-baryon source operators}\label{sec:MB3ptfunc}

In order to investigate interactions of P-wave $I=3/2$$N\pi$ and $I=0$ $\Xi\bar{K}$ systems in the HAL QCD method at low energy region, we use the following 4-point correlation functions, 
\begin{eqnarray}
F^{N\pi}_{\alpha,j_{z}}({\bf r},t) 
&=& \bra{0} \pi^{+}({\bf r+x},t) p_{\alpha}({\bf x},t) \ \bar{\Delta}^{++}_{j_{z}}(t_{0})\ket{0}, \label{eq:def_of_Npi3pt}\\
F^{\Xi\bar{K}}_{\alpha,j_{z}}({\bf r},t) 
&=& \bra{0} \frac{1}{\sqrt{2}}(K^{-}({\bf r+x},t) \Xi^{0}_{\alpha}({\bf x},t)-\bar{K}^{0}({\bf r+x},t) \Xi^{-}_{\alpha}({\bf x},t)) \  \bar{\Omega}^{-}_{j_{z}}(t_{0})\ket{0}, \label{eq:def_of_XiKbar3pt}
\end{eqnarray}
where $\bar{\Delta}^{++}_{j_{z}}(t_{0})$ and $\bar{\Omega}^{-}_{j_{z}}(t_{0})$ are the 3-quark type source operators corresponding to $\Delta$ and $\Omega$ baryons, respectively. 

\begin{figure}
    \centering
    \includegraphics[width=1.0\textwidth]{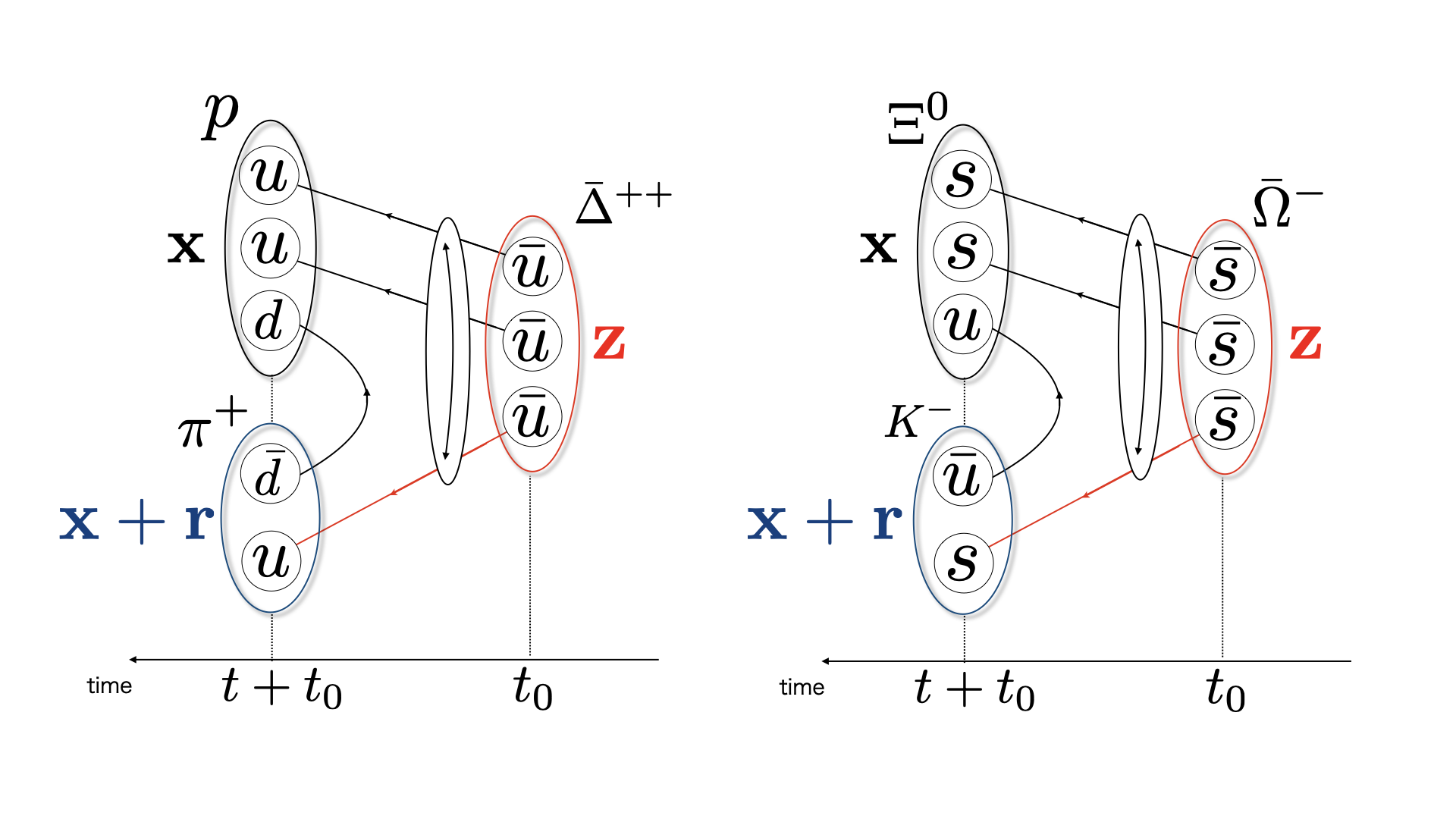}
    \caption{Quark contraction diagram for Eq.~(\ref{eq:def_of_Npi3pt}) (Left) and Eq.~(\ref{eq:def_of_XiKbar3pt}) (Right). A circles with two-way arrows across three lines means a permutation of quark contractions among them. All-to-all propagators are used in red lines.}
    \label{fig:contraction}
\end{figure}

We show quark contraction diagram for Eq.~(\ref{eq:def_of_Npi3pt}) and that for the first term in Eq.~(\ref{eq:def_of_XiKbar3pt}) in Fig.~\ref{fig:contraction}. The second term in Eq.~(\ref{eq:def_of_XiKbar3pt}) is the same as the first one under the isospin symmetry for $m_{u}=m_{d}$.
Since we fix ${\bf x}$ but sum over ${\bf z}$ and we need the dependence on all ${\bf r}$, quark propagators represented by red lines in this figure must be all-to-all propagators. We use the conventional stochastic technique to estimate them approximately, together with the dilution~\cite{Foley:2005ac} for color/spinor components and the s2 dilution~\cite{Akahoshi:2019klc}.

\subsection{Simulation details}\label{sec:simdetails}

In our numerical calculations, we use (2+1)-flavor gauge configurations generated by PACS-CS Collaboration with the improved Iwasaki gauge action and the $\order{a}$-improved Wilson quark action at $\beta = 1.90$ on $32^3 \times 64$ lattice~\cite{PACS-CS:2008bkb}, which corresponds to $a\approx0.09$~fm for the lattice spacing. The hopping parameters of the ensemble in our calculations are $\kappa_{u(d)}=0.13754$ and $\kappa_{s}=0.13640$. The periodic boundary condition is imposed in all spacetime directions. We used 450 configurations with 16 sources at different time slices on each configuration. Statistical errors are estimated by the jackknife method with a binsize of 45 configurations.

We employ the smeared quark source using the Gaussian-type smearing function~\cite{Iritani:2016jie}, where we take $(A,B)=(1.2, 0.17)$ for light quarks and $(A,B)=(1.2, 0.25)$ for strange quarks in lattice unit. We also apply the same smearing to the quarks at the sink with $(A,B)=(1.0, 1/0.7)$ to reduce singular behaviors of the potentials at short distances~\cite{Akahoshi:2021sxc}, which are difficult to fit.

In order to reduce statistical fluctuations, we use the truncated solver method~\cite{Bali:2009hu} combined with the covariant-approximation averaging~\cite{Shintani:2014vja}, namely all-mode averaging without low modes. In this method, we use the translational invariance of the baryon operators at the sink. We take 64 different spatial points in our calculation, given by ${\bf x}+\Delta{\bf x}$ with $\Delta{\bf x} = (0,0,0), (0,0,8), \cdots, (24,24,24)$.

Masses of several hadrons estimated from 2-point functions in this setup are listed in Table \ref{tab:hadronmass}. Since $\Delta$ and $\Omega$ masses lies below $N\pi$ and $\Xi\bar{K}$ threshold energies, respectively, they appear as bound states in this setup.

\begin{table}[b]
\caption{\label{tab:hadronmass}%
Hadron masses in MeV units estimated by fitting 2-point functions. The second row shows fitting ranges in lattice unit.}
\begin{tabular}{c|cccccc}
hadron&
$\pi$ &
$K$ &
$N$ &
$\Xi$ &
$\Delta$ &
$\Omega$ 
 \\ \hline \hline
mass&  $411.2(1.7)$& $635.1(1.5)$& 
           $1217.2(4.7)$& $1505.3(4.5)$&
            $1522.9(7.8)$& $1847.0(6.5)$
\\ 
fit range &  $[10,30]$ & $[10,30]$& 
           $[7,20]$& $[7,20]$&
            $[6,15]$& $[6,20]$
\end{tabular}
\end{table}

To obtain the NBS wave functions with $J^{P}=3/2^{+}$, we project meson-baryon operators at the sink onto the same component in $H_{g}$ representation of the cubic group as that of the source operators. 
Since the different component provides the same LO potentials, we take an average over components in addition to spins of $N$ and $\Xi$.

\section{Results}\label{sec:results}
\subsection{Potentials}
\begin{figure}[t]
    \begin{center}
        \includegraphics[width=0.49\textwidth]{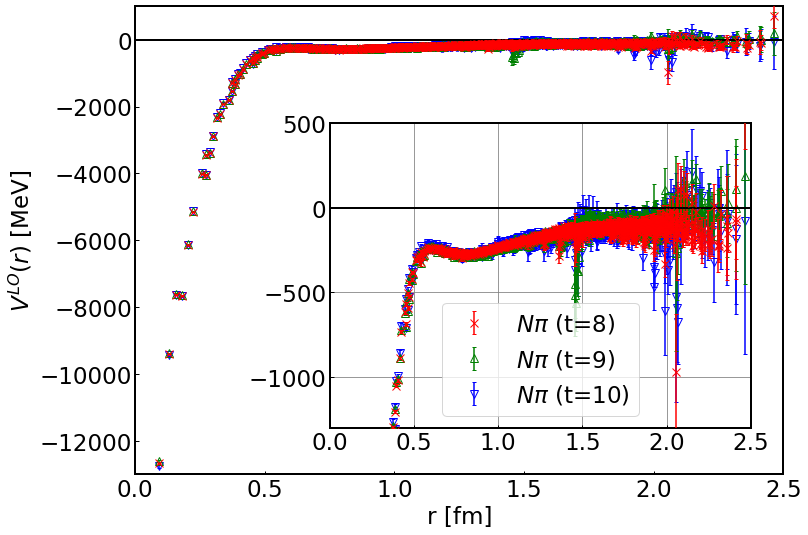}
	    \includegraphics[width=0.48\textwidth]{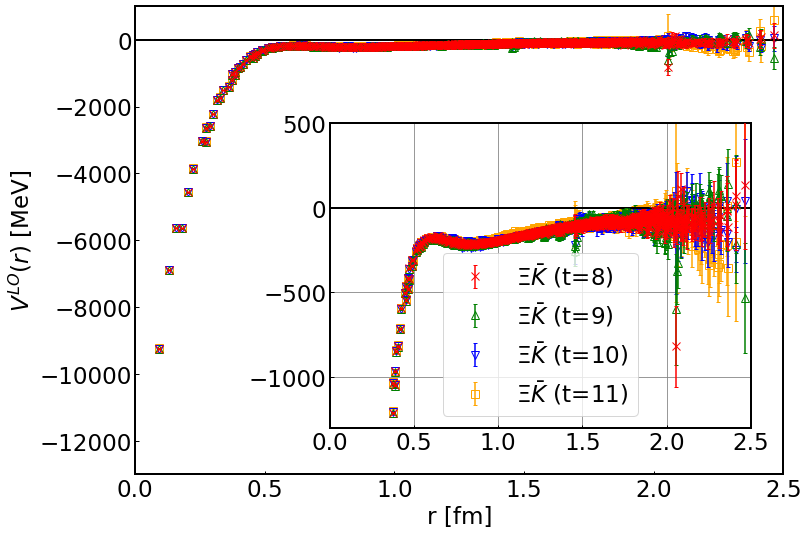}
	    \caption{The leading-order potentials of the $N\pi$ system at $t=8$--$10$ (Left) and the $\Xi\bar{K}$ system at $t=8$--$11$ (Right).}
	    \label{fig:rawpotdata}    
    \end{center}
\end{figure}

In Fig.~\ref{fig:rawpotdata}, we present the LO potentials for P-wave $N\pi$ at $t=8$--$10$ and $\Xi\bar{K}$ at $t=8$--$11$, where we do not see large $t$-dependence of potentials. In addition, the effective mass from each hadron 2-point correlation function is saturated in this time region.

The two potentials have similar behavior to each other. Both have very strong attractions, which are probably needed to produce bound states. More specifically, the $N\pi$ potential has slightly deeper than the $\Xi\bar K$. For middle and long distances, the meson-baryon interaction is probably explained by a meson exchange between meson and baryon. The $N\pi$ system exchanges $\rho$ meson while the $N\pi$ system exchanges $\rho$ meson and the octet part of $\phi$/$\omega$ mesons. Since these mesons have similar masses for a slightly broken flavor SU(3) symmetry, potentials at middle and long distances also become similar.

At $r \approx 1.4, 2.1$ fm, the potentials have the comb-like structure, which are likely to be caused by contributions from higher partial waves survived in the projection of the meson-baryon operators at the sink to the $H_{g}$ representation~\cite{Miyamoto:2019jjc}, such as $J^{P} = 5/2^{+}, 7/2^{+}$ and so on in the present case.

\subsection{Phase shifts and binding energies}

We solve the Schr\"{o}dinger equation for the angular momentum fixed to $l=1$ with potentials obtained from the fit to the original data, and extract phase shifts for the P-wave scattering.

We estimate systematic errors as follows. Systematic errors associated with the finite volume effect are estimated by using two types of fit potentials. One is the three Gaussians as
\begin{eqnarray}\label{eq:3G}
V^{3G}(\vb{r}) 
= a_{0}e^{-(r/a_{1})^2}+a_{2}e^{-(r/a_{3})^2}+a_{4}e^{-(r/a_{5})^2}, 
\end{eqnarray}
where we assume that $a_{1} < a_{3} < a_{5}$, and the other is the modified fitting function as~\cite{Akahoshi:2020ojo}
\begin{eqnarray}
V^{3G}_{P}(\vb{r})
= V^{3G}(\vb{r}) + \sum_{\vb{n}}V^{3G}(\vb{r}+L\vb{n}),
\end{eqnarray}
where $\vb{n} \in \{ (0,0,\pm1),(0,\pm1,0),(\pm1,0,0) \}$. Systematic errors associated with the finite lattice spacing are estimated by two ways. One is a difference between fits with and without data at $r=a$, and the other is
a difference between the laplacian term in Eq.~(\ref{eq:LOpot_murano}) calculated with 2nd and 4th order accuracies.
Finally, we use the potentials from $t=8$ to $10$ for $N\pi$ and from $t=8$ to $11$ for $\Xi\bar{K}$ to estimate systematic errors associated with $t$-dependences, mainly caused by the leading order truncation for the non-local potential. Central values of physical observables are calculated from potential fitted by $V^{3G}_{P}(\vb{r})$ without data at $r=a$ and with the 2nd-order approximation of the laplacian term at $t=9$ for $N\pi$ and at $t=10$ for $\Xi\bar{K}$, and systematic errors are estimated from potentials for the other combinations.

\begin{figure}[t]
    \begin{center}
        \includegraphics[width=0.49\textwidth]{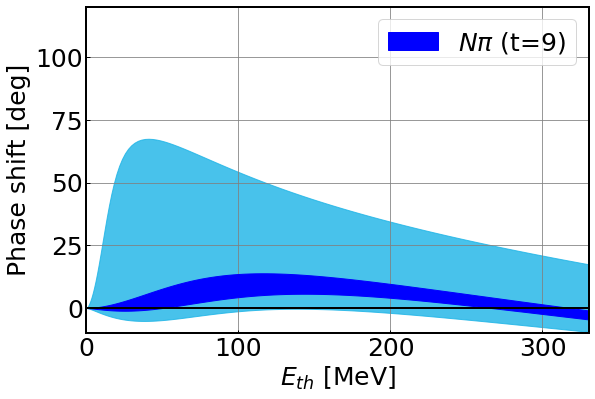}
	    \includegraphics[width=0.48\textwidth]{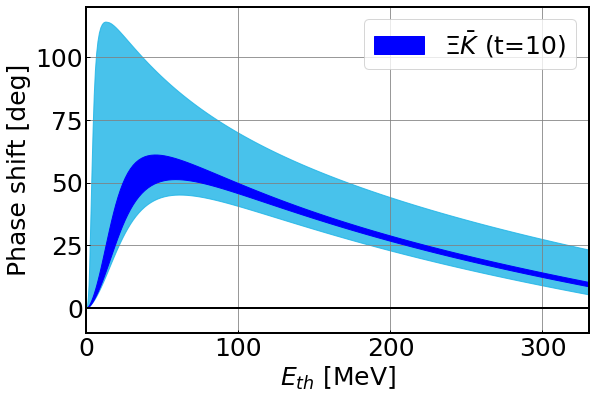}
	    \caption{Scattering phase shifts for $N\pi$ (Left) and $\Xi\bar{K}$ (Right). The dark and the light bands show the statistical and systematic error, respectively.}
	    \label{fig:phase}    
    \end{center}
\end{figure}

Fig.~\ref{fig:phase} shows phase shifts as a function of the energy from the 2-body threshold. Both phase shifts show attractive behaviors, which are consistent with the shape of potentials. We also find no resonances but a bound state in the behavior of phase shifts in both channels. Indeed, phase shifts approach $-180^{\circ}$ when we increase energies, rather than passing 90 degrees.

We calculate binding energies by solving the Schr\"{o}dinger equation via the Gaussian Expansion Method (GEM)~\cite{Hiyama:2003cu}, which gives
\begin{eqnarray}
 E^{N\pi}_{\textrm{bind}} &=& 82.6 (8.2) \smqty(+90.1 \\ -38.4)~\textrm{MeV}, \\
 E^{\Xi\bar{K}}_{\textrm{bind}} &=& 193.9(3.2) \smqty(+163.8 \\ -11.4)~\textrm{MeV},
\end{eqnarray}
where the first and second error represent statistical and systematic errors estimated in the same way as those of the phase shifts, respectively. Both agree with values estimated from the $\Delta$ 2-point function $m_{N}+m_{\pi}-m_{\Delta} = 105.5(5.2)~\textrm{MeV}$ and the $\Omega$ 2-point functions $m_{\Xi}+m_{\bar{K}}-m_{\Omega} = 293.5(2.8)~\textrm{MeV}$ within rather large systematic errors, which come dominantly from the lattice artifact at short distances.

The $\Xi \bar K$ system has a larger binding energy than that of the $N\pi$ by a hundred or more MeV. Since the two system have the similar interactions, it is suggested that a mass difference between $\Omega$ and $\Delta$ comes mainly from a difference in the kinetic energy between the $N\pi$ and $\Xi\bar{K}$ systems. We expect that as we decrease quark masses, $\Delta$ becomes a resonance while $\Omega$ remains to be a bound state due to the small kinetic energy of $\Xi\bar{K}$ due to its large reduced mass.

\section{Conclusion}\label{sec:conclusion}

In this paper, we have analyzed interactions in P-wave $I=3/2$  $N\pi$  and $I=0$ $\Xi\bar{K}$ systems using the HAL QCD method with 3-quark type source operators. 
We have employed gauge configurations with $m_{\pi}\approx 410$~MeV, where the $\Delta$ baryon exists as a bound state in the $N\pi$ system. We solved the Schr\"{o}dinger equation with fitted potentials to extract scattering phase shifts and binding energies. 

A fact that attractive potentials are similar between both systems indicates that the mass difference between $\Delta$ and $\Omega$ baryons mostly comes from the difference of the kinematical structure of $N\pi$ and $\Xi\bar{K}$ systems. The binding energies are consistent with those estimated from the $\Delta$ and $\Omega$ 2-point functions within large errors, mostly due to the lattice artifact at short distances.

In this work, we have performed the LO analysis in the derivative expansion of  non-local potentials in the HAL QCD method. This is enough in this setup because we are interested in the bound state corresponding to the $\Delta$ baryon, which lies in the very low energy region. In order to study the $\Delta$ as a resonance, which is one of our future works, however, we may have to perform the next-leading order (NLO) analysis to extend the energy region above the threshold.

\section{Acknowledgements}
We thank the PACS-CS Collaboration~\cite{PACS-CS:2008bkb} and ILDG/JLDG~\cite{Amagasa:2015zwb} for providing us their gauge configurations. We use lattice QCD code of Bridge++ ~\cite{bridge++url, Ueda:2014rya} and its optimized version for the Oakforest-PACS by Dr. I. Kanamori~\cite{Kanamori:2018hwh}.  Our numerical calculation has been performed on HOKUSAI BigWaterfall at RIKEN and Oakforest-PACS in Joint Center for Advanced HighPerformance Computing (JCAHPC). This work is supported in part by the HPCI System Research Project (Project ID: hp200108(FY2020), hp210061(FY2021)), by Multidisciplinary Cooperative Research Program in CCS, University of Tsukuba, and by the Grant-in-Aid of the MEXT for Scientific Research ( Nos. JP16H03978, JP18H05236). Y.A. is supported in part by the Japan Society for the Promotion of Science (JSPS). We thank other members of the HAL Collaboration for fruitful discussions. We also appreciate Dr. J. Bulava for providing us several projected meson-baryon operators onto the irreps of the cubic group, which are used for our test calculations in this work.

\end{document}